\documentclass[iop]{emulateapj}

\usepackage{amsmath}

\slugcomment{draft \today}

\shorttitle{Small-Scale Dynamo in Primordial Star Formation}
\shortauthors{Schober et al.}

\begin{document}

\title{The Small-Scale Dynamo and Non-Ideal MHD in Primordial Star Formation}

\author{Jennifer Schober\altaffilmark{1},
        Dominik Schleicher\altaffilmark{2}, 
        Christoph Federrath\altaffilmark{3,1}, 
        Simon Glover\altaffilmark{1},
        Ralf S.~Klessen\altaffilmark{1}, 
        Robi Banerjee\altaffilmark{4}}
\email{schober@stud.uni-heidelberg.de}
\email{dschleic@astro.physik.uni-goettingen.de}
\email{christoph.federrath@monash.edu}
\email{glover@uni-heidelberg.de}
\email{klessen@uni-heidelberg.de}
\email{banerjee@hs.uni-hamburg.de}
\altaffiltext{1}{Zentrum f\"ur Astronomie der Universit\"at Heidelberg, Institut f\"ur Theoretische Astrophysik, Albert-Ueberle-Str.~2, D-69120 Heidelberg, Germany}
\altaffiltext{2}{Georg-August-Universit\"at G\"ottingen, Institut f\"ur Astrophysik, Friedrich-Hund-Platz, 37077 G\"ottingen, Germany}
\altaffiltext{3}{Monash Centre for Astrophysics (MoCA), School of Mathematical Sciences, Monash University, Vic 3800, Australia}
\altaffiltext{4}{Hamburger Sternwarte, Gojenbergsweg 112, 21029 Hamburg, Germany}

\begin{abstract}
We study the amplification of magnetic fields during the formation of primordial halos. The turbulence generated by gravitational infall motions during the formation of the first stars and galaxies can amplify magnetic fields very efficiently and on short timescales up to dynamically significant values. Using the Kazantsev theory, which describes the so-called small-scale dynamo - a magnetohydrodynamical process converting kinetic energy from turbulence into magnetic energy - we can then calculate the growth rate of the small-scale magnetic field. Our calculations are based on a detailed chemical network and we include non-ideal magnetohydrodynamical effects such as ambipolar diffusion and Ohmic dissipation. We follow the evolution of the magnetic field up to larger scales until saturation occurs on the Jeans scale. Assuming a weak magnetic seed field generated by the Biermann battery process, both Burgers and Kolmogorov turbulence lead to saturation within a rather small density range. Such fields are likely to become relevant after the formation of a protostellar disk and, thus, could influence the formation of the first stars and galaxies in the Universe.
\end{abstract}


\date{\today}

\maketitle

\section{Introduction}

Magnetic fields play an important role in star formation. There is increasing evidence that dynamically important magnetic fields were also present in the early Universe \citep{BanerjeeJedamzik2004,SilkLanger2006,SchleicherEtAl2010,SurEtAl2010,FederrathEtAl2011b,TurkEtAl2012}. If this turns out to be true, current models for the formation of the first stars and galaxies need to be revisited. \\
The theory of primordial star formation has gone through a change recently. It was previously assumed that the first stars were extremely massive and isolated \citep{AbelEtAl2002,BrommLarson2004}. However, new high-resolution calculations \citep{StacyEtAl2010,ClarkEtAl2011,SmithEtAl2011,GreifEtAl2011.1,GreifEtAl2012} show that the accretion disk of a collapsing primordial halo fragments into multiple stars. The inclusion of magnetic fields could change this picture again. The magnetic pressure can stabilize an accretion disk and, depending on the field strength, suppress the fragmentation \citep{MachidaEtAl2004,HennebelleTeyssier2007,PriceBate2007,PetersEtAl2011,SeifriedEtAl2011}. So far, there are only a few studies on primordial star formation that include magnetic fields \citep{TanBlackman2004,MakiSusa2004,SilkLanger2006,MakiSusa2007,SchleicherEtAl2010,Machida2010,FederrathEtAl2011a,SurEtAl2012,TurkEtAl2012}. The magnetic field is expected to have similar effects to those seen in present-day star formation, such as the launching of winds and jets \citep{MachidaEtAl2006,MachidaEtAl2008a}. The latter eject gas from the accretion disk which could otherwise have collapsed onto the star. Thus, the star formation efficiency is reduced, especially for high-mass stars \citep{TanBlackman2004,Machida2010}. Strong jets can transport matter even out of the star-forming halo, leading to a magnetization of the intergalactic medium \citep{XuEtAl2011}.\\
For developing primordial star formation theory further, we need to know structure and strength of the magnetic fields. The mechanism we suggest for producing strong fields is the small-scale dynamo. This magnetohydrodynamical (MHD) process amplifies weak magnetic seed fields exponentially by converting kinetic energy from the turbulence into magnetic energy.\\
\citet{Kazantsev1968} developed a theory for describing the small-scale or turbulent dynamo. The main equation of this theory is the Kazantsev equation, the eigenvalue of which is the growth rate of the magnetic energy. There are different solutions of this equation \citep{RogachevskiiKleerin1997,KleeorinRogachevskii2011,Schekochihin2002}. In this work we use the solution proposed by \citet{SchoberEtAl2012}, which takes into account different types of turbulence. The growth rates of the magnetic field obtained in that work are comparable to previous results for Kolmogorov turbulence \citep{Subramanian1997,BrandenburgSubramanian2005}.\\
We model the physical and chemical processes during the collapse of a primordial halo to quantitatively determine the properties of the small-scale dynamo. We calculate the magnetic Prandtl number and the magnetic Reynolds number. The latter is compared to the critical magnetic Reynolds number for small-scale dynamo action. Furthermore, we calculate the growth rate, which depends on the magnetic Prandtl number and the hydrodynamic Reynolds number. We assume a weak initial magnetic field of $10^{-20}$ G on the viscous scale produced by the Biermann battery \citep{Biermann1950,KulsrudEtAl1997,Xu2008}. This allows us to determine the evolution of the magnetic field strength during the collapse.\\
The structure of our study is as follows. In Section \ref{Properties} we review the properties of primordial gas. We present our numerical calculation of the chemistry and thermal evolution of the gas and discuss the characteristic magnetohydrodynamical quantities. Furthermore, we discuss the origin of turbulence and weak magnetic seed fields in primordial halos, which are essential for operation of the small-scale dynamo. Section \ref{SmallScale} concentrates on the small-scale magnetic field evolution. We give the basic equations of the Kazantsev theory and the resulting growth rates of the magnetic energy. We apply these results for the small-scale magnetic field to our model for the collapse of a primordial halo. In Section \ref{LargeScale} we present a model for the transport of the magnetic energy to larger scales. This allows us to calculate the magnetic energy on the Jeans scale of the primordial halo.

\section{Properties of the Primordial Gas}
\label{Properties}

\subsection{Chemical and Thermal Evolution}

\begin{figure}
    \centering
    \includegraphics[width=0.45\textwidth]{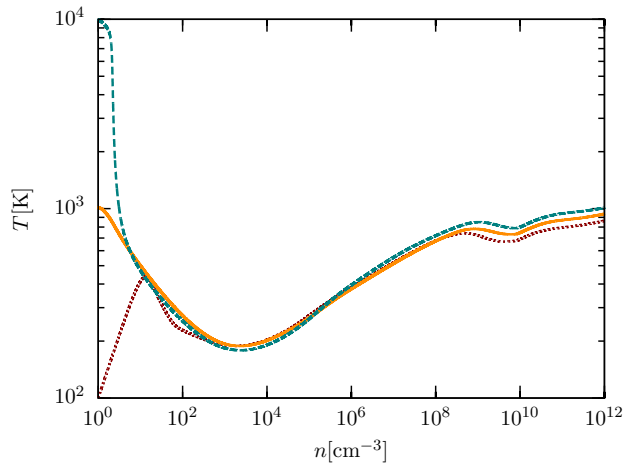}
    \caption{The temperature as a function of the number density. The different lines indicate different initial conditions. The dashed green line corresponds to an initial temperature of $10^4$~K, the dotted red line to $10^3$~K and the solid orange line to $10^2$~K.}
    \label{Figure1}
\end{figure} 
We determine the chemical and thermal evolution of gravitationally collapsing primordial gas using the one-zone model of \citet{GloverSavin2009}, together with a modification implemented by \citet{SchleicherEtAl2009} that relates the collapse time to the equation of state. Moreover, we have included additional Li$^+$ chemistry by using the reaction rates from \citet{BovinoEtAl2011.1} and HeH$^+$ chemistry according to \citet{BovinoEtAl2011.2}. \citet{GloverSavin2009} model the chemistry of the gas with a chemical network that includes around 30 different atomic and molecular species linked by around 400 different chemical reactions. In our calculations, we use the same initial chemical abundances as in the default model in \citet{GloverSavin2009}. The elemental abundances of helium, deuterium and lithium relative to hydrogen are taken to be 0.083 for helium, $2.6 \times 10^{-5}$ for deuterium and $4.3 \times 10^{-10}$ for lithium \citep{Cyburt2004}. The initial density and temperature of the gas were assumed to be $n_{0} = 1 \: {\rm cm^{-3}}$ and $T_{0} = 1000$~K, respectively, but we have verified that our results are not sensitive to these values.\\
In the one-zone model the mass density $\rho$ evolves as
\begin{equation}
    \frac{\text{d} \rho}{\text{d} t} \propto \frac{\rho}{t_\text{ff}},
\end{equation}
where $t_\text{ff} = \sqrt{3\pi/(32 G \rho)}$ is the free-fall time. Moreover, the temperature evolution is determined by the energy equation,
\begin{equation}
    \frac{\text{d} \epsilon}{\text{d} t} = \frac{p}{\rho^2} \frac{\text{d}\rho}{\text{d} t} - \Lambda_\text{cool} + \Lambda_\text{heat},
\end{equation}
\begin{figure}
    \centering
    \includegraphics[width=0.45\textwidth]{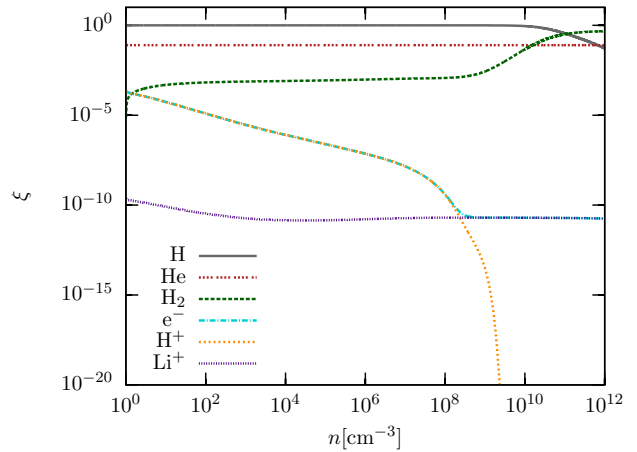}
    \caption{The fractional abundances of different chemical species as a function of the number density.}
    \label{Figure2}
\end{figure} 
where $\epsilon$ is the specific internal energy, $p$ is the thermal pressure and $\Lambda_\text{cool}$ and $\Lambda_\text{heat}$ are the total cooling and the heating rate per unit mass, respectively. The resulting temperature evolution is shown in Figure \ref{Figure1}. Besides the initial temperature of $10^3$~K, we also show in this figure the results for the initial temperatures $10^2$~K and $10^4$~K. Our calculations result in roughly the same evolution for all the initial temperatures after an increase of the density of about one order of magnitude. \\
In Figure \ref{Figure2}, the fractional abundances of H, He, H$_{2}$, H$^{+}$, Li$^{+}$ and free electrons vary with increasing density in our calculations. The abundance of H is constant at low densities, but decreases at densities higher than about $10^{10}~{\rm cm^{-3}}$ due to the formation of H$_{2}$. As there is no dust in primordial gas, large quantities of H$_2$ are produced only at high densities, via the three-body reactions:
\begin{subequations}
  \begin{eqnarray}
  {\rm H} + {\rm H} + {\rm H} & \rightarrow & {\rm H_{2}} + {\rm H}, \\
  {\rm H} + {\rm H} + {\rm He} & \rightarrow & {\rm H_{2}} + {\rm He}, \\
  {\rm H} + {\rm H} + {\rm H_{2}} & \rightarrow & {\rm H_{2}} + {\rm H_{2}}.
  \end{eqnarray}
\end{subequations} 
For the magnetic properties of the primordial gas the abundances of the charged species are especially important. They determine for example the conductivity, which is calculated in the next section. At densities $n < 10^{8} \: {\rm cm^{-3}}$, ionized hydrogen is the main positive ion, while at higher densities, Li$^{+}$ dominates. The sharp drop in the H$^{+}$ abundance at densities $n > 10^{8} \: {\rm cm^{-3}}$ results from the removal of H$^{+}$ from the gas by the reaction chain \citep{GloverSavin2009},
\begin{subequations}
  \begin{eqnarray}
  {\rm H_{2}} + {\rm H^{+}} & \rightarrow & {\rm H_{3}^{+}} + \gamma, \\
  {\rm H_{3}^{+}} + {\rm e^{-}} & \rightarrow & {\rm H_{2}} + {\rm H}.
  \end{eqnarray}
\end{subequations}

\subsection{Characteristic Magnetohydrodynamical Quantities}

\paragraph{Viscosity}
It can be shown that the kinematic viscosity is 
\begin{equation}
	\nu = \frac{1}{4 d^2 n}\left(\frac{k T}{\pi m}\right)^{1/2},
\label{nu}
\end{equation}
if the molecules are assumed to be rigid spheres \citep{Choudhuri1998}. Here, $n=\rho/m$ is the number density, $k$  the Boltzmann constant and $T$ the temperature. Furthermore, $d=\sum_\text{i} \xi_\text{i}d_\text{i}$ is the mean particle diameter and $m=\sum_\text{i} \xi_\text{i}m_\text{i}$ the mean mass. $\xi_\text{i}$ is the relative abundance of the species i, $m_\text{i}$ and $d_\text{i}$ are the masses and the Van-der-Waals diameters, respectively. The temperature as well as the abundances of the individual species are functions of the number density.

\paragraph{Diffusivity}
For calculating the magnetic diffusivity $\eta$ we need the conductivity of the gas. In a plasma the three most important contributions to the conductivity of a neutral species (indicated by index n) are \citep{WardleNg1999},
\begin{subequations}
\begin{eqnarray}
	\sigma_{||,\text{n}} & = & \frac{c}{B} \sum_{\text{s}} \xi_{\text{s}} n q_{\text{s}} \beta_{\text{ns}}, \\
	\sigma_{\text{P},\text{n}} & = & \frac{c}{B} \sum_{\text{s}} \xi_{\text{s}} n q_{\text{s}} \frac{\beta_{\text{ns}}}{1+\beta_{\text{ns}}^2},\\
	\sigma_{\text{H},\text{n}} & = & \frac{c}{B} \sum_{\text{s}} \xi_{\text{s}} n q_{\text{s}} \frac{1}{1+\beta_{\text{ns}}^2},
\end{eqnarray}
\label{conductivities}
\end{subequations}
as given in \citet{PintoEtAl2008}. The Hall-parameters $\beta_{\text{ns}}$ are defined as
\begin{equation}
	\beta_{\text{ns}} = \frac{q_{\text{s}} B}{m_{\text{s}} c} ~ \frac{m_{\text{s}} + m_{\text{n}}}{m_{\text{n}}\xi_{\text{n}} n \left\langle \sigma v \right\rangle_{\text{sn}}}.
\end{equation}
Here, $m_{\text{n}}$ and $m_{\text{s}}$ are the masses of the neutral and the charged particles, $\xi_{\text{n}}$ and $\xi_{\text{s}}$ are the abundance fractions of the species, and $\left\langle \sigma v \right\rangle_{\text{sn}}$ is the momentum transfer rate coefficient. We take these coefficients, which are functions of the temperature, from \citet{PintoGalli2008}, where we use the polarisation approximation for Li$^+$.\\
The two dominant effects that lead to the dissipation of magnetic energy are the Ohmic resistivity and ambipolar diffusion. We can neglect the contribution of the Hall effect to the resistivity, as here the force acts perpendicular to the current and, thus, no energy is dissipated into heat. We calculate the distributions of the Ohmic resistivity and the ambipolar diffusion by
\begin{subequations}
\begin{eqnarray}
	\eta_{\text{Ohm},\text{n}} & = & \frac{c^2}{4 \pi \sigma_{||,\text{n}}},  \\
	\eta_{\text{AD},\text{n}} & = & \frac{c^2}{4 \pi} \left(\frac{\sigma_{\text{P,n}}}{\sigma_{\text{P,n}}^2+ \sigma_{\text{H,n}}^2}-\frac{1}{ \sigma_{||,\text{n}}}\right).
\end{eqnarray}
\label{eta}
\end{subequations}
We focus on the most important neutral species H, He and $\text{H}_2$ and the charged species $\text{H}^+$, $\text{e}^-$ and $\text{Li}^+$. For each neutral species we calculate the resistivities $\eta_{\text{Ohm},\text{n}}$ and $\eta_{\text{AD},\text{n}}$. The magnetic field strength $B$ drops out in the Ohmic case. Finally, the total Ohmic magnetic diffusivity is $\eta_{\text{Ohm}} = \sum_{\text{n}} \eta_{\text{Ohm,n}}$ and the total resistivity due to ambipolar diffusion is $\eta_{\text{AD}} = 1/(\sum_{\text{n}} \eta_{\text{AD,n}}^{-1})$.\footnote{From private communication with Daniele Galli.}

\begin{figure}
    \centering
    \includegraphics[width=0.45\textwidth]{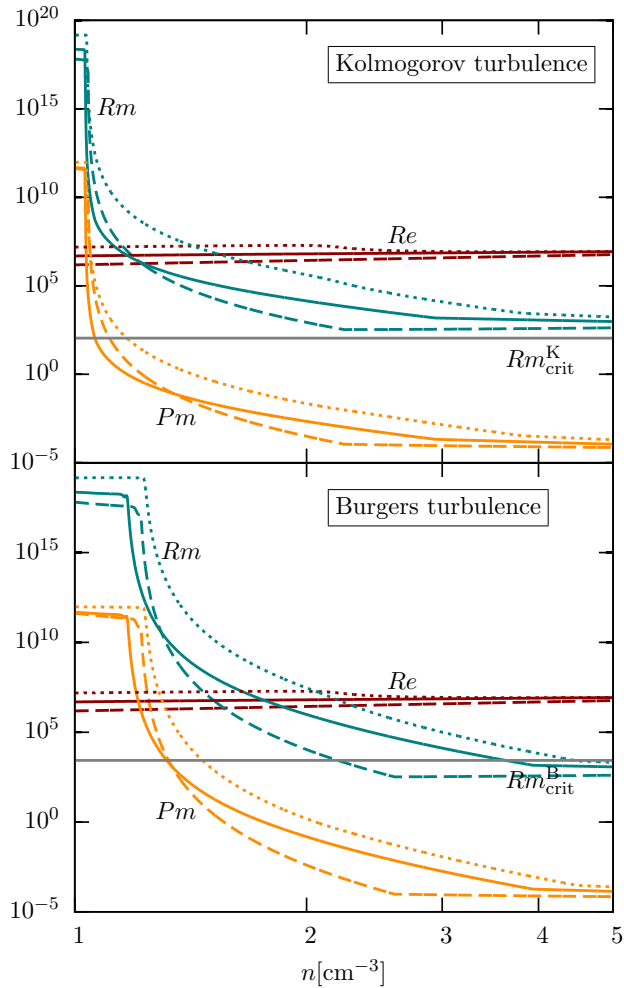}
    \caption{The hydrodynamic and magnetic Reynolds numbers, $Re$ and $Rm$, as well as the magnetic Prandtl numbers, $Pm$, on the current Jeans scale. The numbers are presented as a function of the hydrogen nuclei number density $n$. The solid lines represent an initial temperature of $10^3$ K, the dashed lines $10^2$ K and the dotted lines $10^4$ K. Moreover, the horizontal lines indicate the critical magnetic Reynolds number for Kolmogorov and Burgers turbulence ($Rm_\text{crit}^\text{K}=107$ and $Rm_\text{crit}^\text{B}=2718$) as derived in \citet{SchoberEtAl2012}. The rapid decrease of $Rm$ and $Pm$ from the very high starting values is caused by the exponential dynamo amplification of the magnetic field. We show the results for Kolmogorov turbulence in the upper plot and the results for Burgers turbulence in the lower plot.}
    \label{Figure3}
\end{figure}

\paragraph{Reynolds Numbers}
The hydrodynamic and magnetic Reynolds numbers are defined as 
\begin{subequations}
\begin{eqnarray}
	Re & \equiv & \frac{VL}{\nu} \\
	Rm & \equiv & \frac{VL}{\eta},
\label{Reynolds}
\end{eqnarray}
\end{subequations}
where $L$ is the length of the largest turbulent fluctuations and $V$ the typical velocity on that scale. Notice, that we give these numbers on the forcing scale, i.e.\ the Jeans scale, which means $L = \ell_\text{J}$ and $V = v_\text{J}$. \\
For the calculation of the magnetic Reynolds number we use the sum of $\eta_\text{Ohm}$ and $\eta_\text{AD}$. The resulting Reynolds numbers are shown in Figure \ref{Figure3} as a function of the density. The critical magnetic Reynolds numbers (\ref{Rmcrit}) are also indicated for the two extreme types of turbulence. 

\paragraph{Magnetic Prandtl Number}
The definition of the magnetic Prandtl number is 
\begin{equation}
  Pm \equiv \frac{Rm}{Re} = \frac{\nu}{\eta}. 
\label{Pm}
\end{equation}
We can calculate this quantity by using the equations (\ref{nu}) and (\ref{eta}). In Figure \ref{Figure3} the density dependency of the magnetic Prandtl number is shown for both Kolmogorov and Burgers turbulence. For clarification we point out that the rapid decrease of the magnetic Reynolds and Prandtl number is caused by the dynamo amplification of the magnetic field. In the beginning of the collapse Ohmic resistivity is the dominant diffusion process. With increasing magnetic field $\eta_\text{AD}$ increases proportional to $B^2$ (see equations \ref{conductivities} to \ref{eta}) and becomes the main process for magnetic diffusion. Since $Rm$ and $Pm$ are both proportional to $1 / \eta_{\rm AD}$, in the limit where $\eta_{\rm AD} \gg \eta_{\rm Ohm}$, both decrease rapidly with increasing magnetic field strength.\\
In addition, we tested the influence of varying the initial temperature on the evolution of the Reynolds numbers and the magnetic Prandtl number. For the initial temperature ranging from $10^2$ K to $10^4$ K we found only small variations in the Reynolds numbers and the magnetic Prandtl number, as illustrated in Figure \ref{Figure3}.

\subsection{Turbulence}

Turbulence is an omnipresent phenomenon in astrophysics. Turbulence is observed, for example, in the convection zone of stars, in accretion discs, and in the interstellar medium.\\
For the first star-forming halos considered here, we assume that turbulence is driven by virialisation \citep{WiseAbel2007,GreifEtAl2008} or by accretion of gas into the center of the halos \citep{KlessenHennebelle2010,ElmegreenBurkert2010,FederrathEtAl2011a}. The presence of turbulence affects star formation strongly, as the turbulent pressure works against the collapse to a star \citep{VazquezEtAl1998,MacLowKlessen2004,KrumholzMcKee2005,McKeeOstriker2007}. Moreover, \citet{FederrathEtAl2011a} show in a Fourier analysis that the turbulence is effectively driven on the Jeans scale.\\
There are different types of turbulence. In this paper we concentrate on the two extreme cases, Kolmogorov turbulence and highly compressible Burgers turbulence. The different types are described in the inertial range by the relation between the length scale $\ell$ and the velocity $v$ on that scale,
\begin{equation}
  v \propto \ell^{\vartheta}.
\end{equation}
The exponent $\vartheta$ ranges from 1/3 for incompressible turbulence \citep{Kolmogorov1941} to 1/2 for highly compressible turbulence \citep{Burgers1948}. In real astrophysical objects we expect the turbulence index $\vartheta$ to lie between these extreme cases \citep{KritsukEtAl2007,SchmidtEtAl2008,FederrathEtAl2010}.

\subsection{Magnetic Seed Fields}

There are different theories that describe the origin of weak primordial magnetic fields. The first seed fields could already have been produced during inflation. \citet{TurnerWidrow1988} find that a magnetic field $B_0\approx10^{-25}$--$10^{-1}$ nG on a scale of $1$ Mpc can be produced when the conformal invariance is broken. \\
Following \citet{Sigl1997}, there is also a possibility to create a magnetic field during first-order phase transitions in the very early Universe. They predict a field strength $B_0 \approx 10^{-20} ~\text{nG}$ from the electroweak phase transition and $B_0 \approx 10^{-11} ~\text{nG}$ from the QCD phase transition on a scale of $10$ Mpc. \\
Another popular way to generate magnetic fields are battery mechanisms. The so-called Biermann battery uses the fact that electrons and ions have very different masses. If there is a pressure gradient in the plasma the particles get accelerated. Due to their smaller masses the electrons are more strongly accelerated than the ions. This leads to charge separation and an electric field is generated. If the electron density $n_\text{e}$ is constant in space the electric field is static, however, if it varies in space electric currents are generated, which give rise to a corresponding magnetic field. Note, that for the Biermann term to be non-zero the cross product of the electron pressure gradient and the electron density gradient needs to be non-zero \citep{Biermann1950,KulsrudZweibel2008}.

\section{Magnetic Field Amplification on the Viscous Scale}
\label{SmallScale} 

In this section we analyse the evolution of the small-scale magnetic field. We outline the Kazantsev theory, which gives the growth rates of the magnetic field on the viscous scale. Together with the amplification due to gravitational compression and dissipation processes we can calculate the resulting small-scale magnetic field evolution.

\subsection{Small-Scale Dynamo Growth}

\paragraph{Kazantsev Theory}
An arbitrary magnetic field $\textbf{B}$ can, in general, be separated into a mean component $\textbf{B}_0$ and a fluctuating component $\delta \textbf{B}$ with 
\begin{equation}
  \textbf{B} = \textbf{B}_0 + \delta \textbf{B}.
\label{B0deltaB}
\end{equation} 
The induction equation,
\begin{equation}
  \frac{\partial \textbf{B}}{\partial t} = \nabla\times\left(\textbf{v}\times\textbf{B} - \eta\nabla\times\textbf{B}\right),
\label{induction}
\end{equation} 
describes the time evolution of this field, where $\textbf{v}$ is the velocity and $\eta$ the magnetic diffusivity. Substitution of (\ref{B0deltaB}) into the induction equation leads to two equations: an equation for the large-scale field evolution and the Kazantsev equation \citep{Kazantsev1968,BrandenburgSubramanian2005}, which describes the small-scale evolution of the field. \\
The derivation of the Kazantsev equation is based on the assumption that the fluctuations of the magnetic field as well as the fluctuations of the velocity field are homogeneous and isotropic even if the mean fields are not isotropic. Furthermore, the fluctuations are assumed to be a Gaussian random field with zero mean and the velocity fluctuations are assumed to be delta-correlated in time. For the simplicity helicity of the magnetic field is neglected. With these assumptions the Kazantsev equation is
\begin{equation}
  -\kappa_\text{diff}(r)\frac{\text{d}^2\psi(r)}{\text{d}^2r} + U(r)\psi(r) = -\Gamma \psi(r).
\label{Kazantsev}
\end{equation}
The eigenfunctions of this equation are related to the longitudinal correlation function of the magnetic fluctuations $M_L(r,t)$ by $M_\text{L} \equiv 1/(r^2\sqrt{\kappa_\text{diff}})\psi(r)\text{e}^{2\Gamma t}$. We call $\Gamma$ the growth rate of the small-scale magnetic field. The function $\kappa_\text{diff}$ is the magnetic diffusion coefficient, which contains besides the magnetic diffusivity $\eta$ also a scale-dependent turbulent diffusivity. $U$ is called the ``potential'' of the Kazantsev equation. Both $\kappa_\text{diff}$ and $U$ depend only on the correlation function of the turbulent velocity field and the magnetic diffusivity. The correlation function of the turbulent velocity field in turn depends on the different types of turbulence.\\
With a model for the turbulent correlation function, \citet{SchoberEtAl2012} solved the Kazantsev equation (\ref{Kazantsev}) with the WKB-approximation. They found that the critical magnetic Reynolds number for dynamo action $Rm_\text{crit}$ increases with compressibility. The values that $Rm$ needs to exceed are
\begin{subequations}
\begin{eqnarray}
  Rm_\text{crit}^\text{K} & \approx & 107, \\
  Rm_\text{crit}^\text{B} & \approx & 2718,
\end{eqnarray}
\label{Rmcrit}
\end{subequations}
for Kolmogorov and Burgers turbulence, respectively.\\
Moreover, \citet{SchoberEtAl2012} found different growth rates of the magnetic field for different turbulence models, with
\begin{equation}
  \Gamma = \frac{(163-304\vartheta)}{60}\frac{V}{L}Re^{(1-\vartheta)/(1+\vartheta)}
\label{Gamma}
\end{equation}
in the limit of infinite magnetic Prandtl numbers. Here $V$ is the typical velocity on the largest scale of the turbulent eddies of size $L$ and $Re$ is the hydrodynamical Reynolds number.\\
In this paper we analyse the two extreme types of turbulence, Kolmogorov with $\vartheta=1/3$ and Burgers turbulence with $\vartheta=1/2$. We find in the limit of large magnetic Prandtl numbers
\begin{subequations}
\begin{eqnarray}
  \Gamma^\text{K} & = & \frac{37}{36}\frac{V}{L}Re^{1/2}, \\
  \Gamma^\text{B} & = & \frac{11}{60}\frac{V}{L}Re^{1/3}.
\end{eqnarray}
\label{Gammas}
\end{subequations}
\begin{figure}
    \centering
    \includegraphics[width=0.45\textwidth]{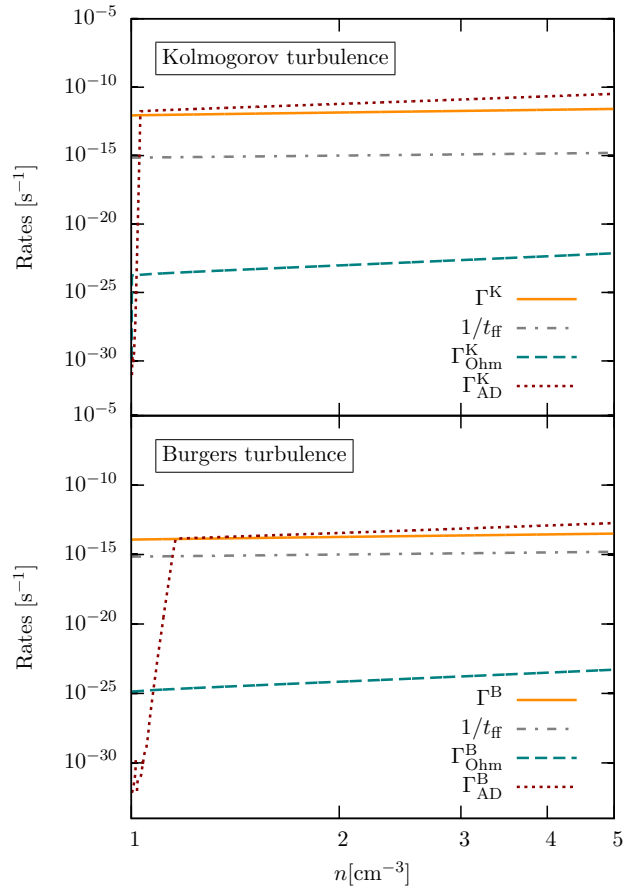}
    \caption{Different characteristic rates on the viscous scale as a function of the density. We compare the growth rate of the magnetic field (solid orange line) to the inverse free fall time (dashed-dotted gray line) and the dissipation rates (Ohmic dissipation: dashed green line, ambipolar diffusion: dotted red line). We show the results for Kolmogorov turbulence in the upper plot and the results for Burgers turbulence in the lower plot.}
    \label{Figure4}
\end{figure}
For the typical velocity of the largest fluctuations we use the sound speed $V = v_\text{J} \approx \sqrt{\gamma k T/m}$, as the Mach number in a primordial halo is roughly one \citep{GreifEtAl2008}. Here $\gamma$ is the adiabatic index. We take $L$ to be the Jeans length, as this is the effective driving scale for turbulence in a collapsing system \citep{SchleicherEtAl2010,FederrathEtAl2011a}. Hence we set $L \approx \ell_\text{J} = \sqrt{\gamma k T/(G m^2 n)}$, where $G$ is the gravitational constant.\\
We compare the growth rate of the small-scale dynamo $\Gamma$ on the viscous scale $\ell_{\nu} = l_\text{J} Re^{-1/(1+\vartheta)}$ to the inverse free-fall time $1/t_{\text{ff}} = [3\pi/(32 G m n)]^{-1/2}$. The result is shown in Figure \ref{Figure4}. In our model the magnetic field on the fastest growing scale increases one to three orders of magnitude faster than the halo collapses. Note that the dynamo growth is exponential in time.

\paragraph{Gravitational Compression}
The gravitational compression due to the collapse of the halo provides additional amplification of the magnetic field.  As long as the condition of flux freezing is fulfilled, the magnetic field $B$ increases with density like 
\begin{equation}
  B \propto n^{2/3}
\end{equation}
for spherically symmetric collapse. Before the dynamo saturates, the amplification by gravitational compression is minor compared to the dynamo growth.

\paragraph{Dissipation}
Part of the magnetic energy is converted into heat by dissipation processes. The dissipation term in the induction equation (\ref{induction}) is $\eta\nabla^2\textbf{B}$. We consider Ohmic dissipation and ambipolar diffusion and approximate this by $\eta B/\ell^2$ and $\partial B/ \partial t$ by $B\Gamma_\text{Ohm}$ and $B\Gamma_\text{AD}$ respectively. We get
\begin{subequations}
\begin{eqnarray}
  \Gamma_\text{Ohm} & \approx & \frac{\eta_\text{Ohm}}{\ell^2},  \\
  \Gamma_\text{AD} & \approx & \frac{\eta_\text{AD}}{\ell^2}.
\end{eqnarray}
\end{subequations}
$\Gamma_\text{Ohm}$ and $\Gamma_\text{AD}$ are the rates of magnetic energy dissipation by Ohmic resistivity and ambipolar diffusion.

\subsection{Critical Magnetic Reynolds Number}

The dependency of the magnetic Reynolds number $Rm$ on the number density is shown in Figure \ref{Figure3} for the two extreme types of turbulence. We also indicate the critical magnetic Reynolds number for small-scale dynamo action $Rm_\text{crit}$. One can see that the magnetic Reynolds number is larger than $Rm_\text{crit}$ at the onset of the collapse. For densities above roughly 4 cm$^{-3}$ $Rm$ becomes smaller than the critical value in the case of Burgers turbulence. For Kolmogorov turbulence $Rm$ becomes also smaller than $Rm_\text{crit}$ for high densities, which are not shown in Figure \ref{Figure3}. However, as we see below, at this point the dynamo is already saturated on the small scale as well as on the large scale. Thus, in the density regimes where the small-scale dynamo operates, the condition $Rm > Rm_\text{crit}$ is always fulfilled.

\subsection{Resulting Small-Scale Magnetic Field}

In principle, the magnetic energy density, $E_\text{B} = B^2/(8\pi)$, evolves as 
\begin{equation}
  \frac{\text{d} E_\text{B}}{\text{d} t} = \left[\Gamma + \frac{4}{3 n} \frac{\text{d} n}{\text{d} t} - \Gamma_\text{Ohm} - \Gamma_\text{AD}(E_\text{B})\right] E_\text{B},
\label{BGrow}
\end{equation}
where we assume spherically symmetric collapse. \\
By solving equation (\ref{BGrow}) numerically we find the evolution of the magnetic energy density on small scales. In Figure \ref{Figure7} we show the resulting growth of the magnetic field strength. As an initial field strength $B_0$ we use $10^{-20}~\text{G}$ on the viscous scale, which is a conservative value for a field generated by a Biermann battery \citep{Biermann1950,Xu2008}. The field strength grows extremely rapidly as the density increases. However, we cannot trust the whole evolution of the magnetic field exactly as shown in Figure \ref{Figure7}. When the field has become strong enough, the magnetic Prandtl number becomes unity or less (see Figure \ref{Figure3}). Then the WKB-approximation breaks down and equations (\ref{Gammas}) are no longer valid. Complementary studies have shown, however, that the small-scale dynamo still operates for $Pm<1$ \citep{BoldyrevCattaneo2004,SchekochihinEtAl2005,SchekochihinEtAl2007,Eyink2011}, although the growth rate may decrease by a factor of a few. We note that \citet{BoldyrevCattaneo2004} find in their studies that the critical magnetic Reynolds number increases with decreasing magnetic Prandtl number. Furthermore, we see in Figure \ref{Figure4} that the ambipolar diffusion rate becomes higher than the growth rate of the magnetic field. In this regime, equation (\ref{Gamma}) is no longer a solution of (\ref{Kazantsev}). We expect that the field grows at the rate (\ref{Gamma}) almost until saturation, but then decreases and the field reaches saturation more slowly.

\subsection{Validity of our Approximation}

In Figure \ref{Figure3} the magnetic Prandtl number $Pm$ is shown as a function of the density. $Pm$ starts with an extremely high value of roughly $10^{12}$ and then after a rather constant phase decreases rapidly.  The magnetic Prandtl number is defined in equation (\ref{Pm}) with $\eta = \eta_\text{Ohm} + \eta_\text{AD}$. For low densities the Ohmic resistivity dominates, which is independent of the magnetic field strength. With increasing density the magnetic field increases due to the dynamo amplification and with $\eta_\text{AD}\propto B^2$ the ambipolar diffusion rate becomes dominant. In this regime the magnetic Prandtl number decreases proportional to $B^{-2}$. As the magnetic field increases exponentially during the small-scale dynamo amplification in the beginning of the collapse, the magnetic Prandtl number decreases rapidly.\\
The approximation of large magnetic Prandtl numbers \citep{SchoberEtAl2012} is accurate during most of the dynamo growth. At the end of the dynamo phase, however, $Pm$ reaches unity and decreases even further and our approximations eventually break down. \citet{SchoberEtAl2012} show that for decreasing $Pm$ the growth rate decreases. However, they make no prediction for the regime $Pm \approx 1$. But numerical simulations show that the dynamo operates also in this regime \citep[e.g.][]{FederrathEtAl2011a}. For $Pm \ll 1$ there is again analytical evidence for small-scale dynamo action \citep[e.g.][]{SchekochihinEtAl2007}. We note that this treatment concerns the viscous scale only.

\section{Magnetic Field Amplification on Larger Scales} 
\label{LargeScale}

\subsection{Model for the Transport of Magnetic Energy to Larger Scales}

\begin{figure}
    \centering
    \includegraphics[width=0.45\textwidth]{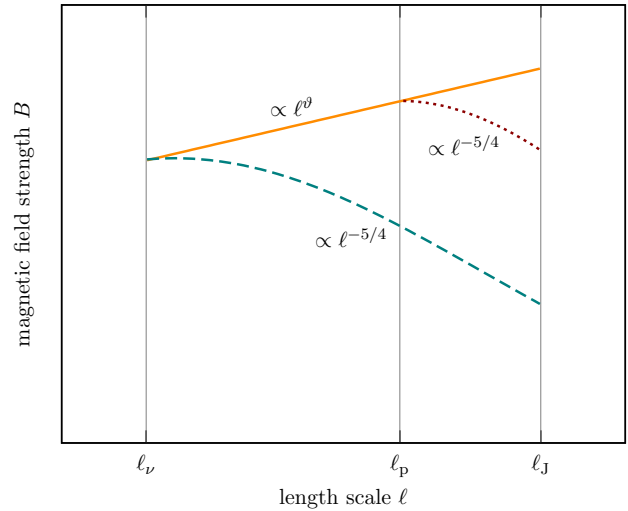}
    \caption{Schematic model for the time evolution of the spectrum of the magnetic field strength in the inertial range of turbulence. For simplicity we use a  fixed frame of reference, where the viscous and the Jeans scale stay constant. The different colors and line types represent the spectrum at different times. The green dashed line shows the spectrum at saturation on the viscous scale. To larger scales the spectrum decreases according to the Kazantsev slope ($B \propto \ell^{-5/4}$). At a time after saturation on the viscous scale the spectrum is indicated by the dotted red line. Finally, the solid orange line shows the spectrum, when the field is saturated on the Jeans scale.}
    \label{Figure5}
\end{figure}
After the magnetic field saturates on the viscous scale the peak of the magnetic energy spectrum moves to larger scales. In this section we present a model for the time evolution of the magnetic energy spectrum. The situation is illustrated schematically in Figure \ref{Figure5}. Here we indicate three different curves, which represent different times. The dashed green line is the spectrum at the time of saturation on the viscous scale, the dotted orange line shows a later time and the solid red line represents an even later point in time at which the magnetic field has saturated on the Jeans scale.\\
During saturation, the coherence length of the magnetic field shifts towards larger scales, a well-known behavior for the small-scale dynamo \citep{SchekochihinEtAl2002,BrandenburgSubramanian2005}, recently shown to be true also in a collapsing system \citep{SurEtAl2012}. Analytical arguments suggest this occurs on the eddy-timescale of the current peak scale $\ell_\text{p}$
\begin{equation}
  \frac{\ell_\text{p}}{v_{\text{p}}} = \frac{\ell_\text{J}}{v_\text{J}} \left(\frac{\ell_\text{p}}{\ell_\text{J}}\right)^{1-\vartheta},
\end{equation}
where we used $v_{\text{p}}=v_\text{J}(\ell_\text{p}/\ell_\text{J})^\vartheta$. Considering that the peak scale moves from the viscous scale $\ell_\nu(t_\nu)$ towards larger scales, we find for the time-dependency of the peak scale
\begin{equation}
  \ell_\text{p}(t) = \ell_\nu(t_\nu) + \left(\frac{v_\text{J}}{\ell_\text{J}^\vartheta} \left(t-t_{\nu}\right)\right)^{1/(1-\vartheta)},
\label{Shift}
\end{equation}
where $t_\nu$ is the point in time, when saturation occurs on the viscous scale.
\begin{figure}
  \centering
  \includegraphics[width=0.45\textwidth]{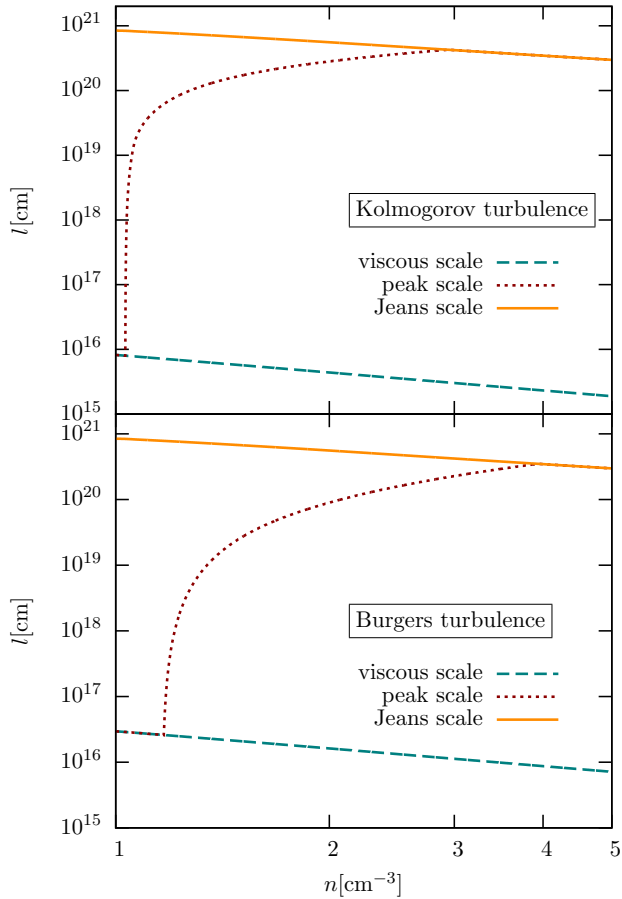}
  \caption{Different characteristic scales as a function of the density. The dashed green line indicates the viscous scale, the dotted red line the scale corresponding to the peak of the magnetic energy spectrum, and the solid orange line the Jeans scale. We show the results for Kolmogorov turbulence in the upper plot and the results for Burgers turbulence in the lower plot.}
    \label{Figure6}
\end{figure}\\
The slope of the curves proportional to $\ell^{-5/4}$ is known as the Kazantsev slope in real space\footnote{In many references the magnetic energy spectrum is given as a function of the wave number $\textit{k}$, defined for example as $B^2/(8\pi\rho) = 1/2 \int M(\textit{k}) \text{d} \textit{k}$. In this case the Kazantsev slope is $M(\textit{k}) \propto \textit{k}^{3/2}$. From this we find $B^2 \propto \textit{k}^{5/2}$ and $B \propto \ell^{-5/4}$.}, which can be derived from the Fourier-transformed Kazantsev equation (\ref{Kazantsev}) \citep{BrandenburgSubramanian2005}. This characteristic slope is also observed in simulations \citep{FederrathEtAl2011a,XuEtAl2011}. The curve that connects the peak maxima at different times (red-colored curve) is a relic of the turbulence spectrum and thus is proportional to $\ell^{\vartheta}$.\\
At each time step we calculate the peak magnetic field strength by solving the stationary case of equation (\ref{BGrow}). However, we find that the magnetic field strength exceeds the equipartition field strength on scales larger than the viscous scale. The reason for this is that the ambipolar dissipation rate, which is proportional to $B^2/\ell^2$, decreases rapidly in this regime. Thus, it cannot balance the growth rate any longer and we need to set the equipartition field strength as an upper limit $B_{\ell,\text{max}}$. With $B_{\ell,\text{max}}^2/(8\pi) = 1/2 \rho v(\ell)^2$ we find the maximum magnetic field strength $B_{\ell,\text{max}} = \sqrt{4 \pi \rho} v(\ell)$. 

Taking the typical turbulent velocity on the scale of the turbulence $\ell$ to be related to the sound-speed by $v(\ell) = (\ell / \ell_J)^\vartheta c_s \simeq (\gamma k T / m)^{1/2} (\ell / \ell_J)^\vartheta$, we find that
\begin{equation}
  B_{\ell,\text{max}} = \sqrt{4 \pi \gamma k T n}~\left(\ell/\ell_\text{J}\right)^{\vartheta}.
\label{Bmax}
\end{equation}
Using the Kazantsev slope, we can extrapolate the magnetic field strength onto the current Jeans length. By this we are able to determine the time evolution of the magnetic field on the Jeans scale.\\
For this process to be relevant during collapse, the eddy-timescale needs to be smaller than the collapse timescale. Thus, the small-scale dynamo is unlikely to produce magnetic fields on scales larger than the Jeans scale. Figure \ref{Figure6} shows the viscous, the peak, and the Jeans scale as a function of density. During the small-scale dynamo growth the spectrum of the magnetic energy peaks at the viscous scale. After saturation on the viscous scale the peak moves to larger scales according to equation (\ref{Shift}) until it reaches the Jeans scale.

\subsection{Resulting Jeans-Scale Magnetic Field}

\begin{figure}
    \centering
    \includegraphics[width=0.45\textwidth]{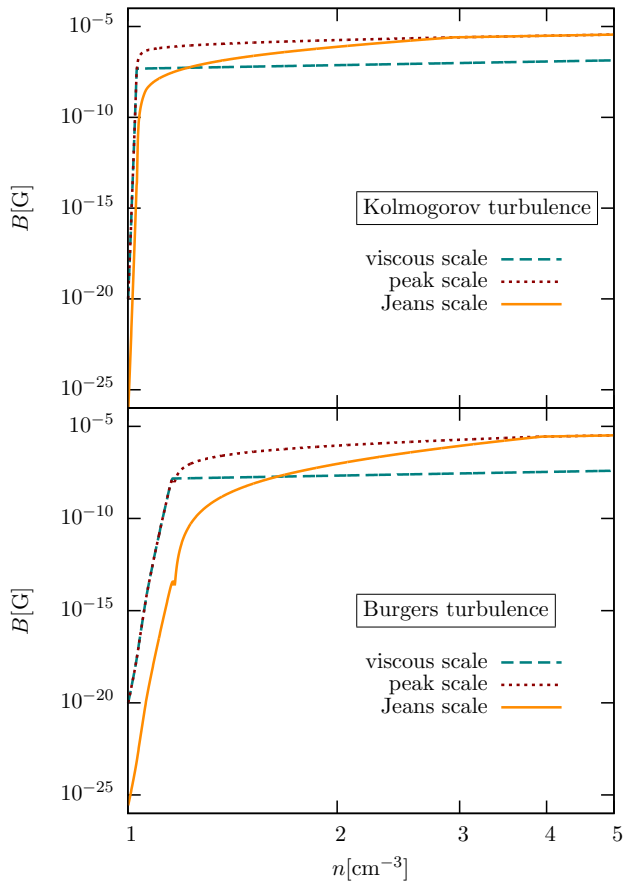}
    \caption{The magnetic field strength as a function of the number density on different scales. The dashed green line corresponds to the field evolution on the viscous scale, the dotted red line to the peak scale and the solid orange line to the Jeans scale. We show the results for Kolmogorov turbulence in the upper plot and the results for Burgers turbulence in the lower plot.}
    \label{Figure7}
\end{figure}
As described in the last section, we determine the magnetic field on the Jeans scale by extrapolation from the peak scale. The result of the large-scale magnetic field is shown in Figure \ref{Figure7} together with the field on the current peak scale and the one on the viscous scale. One can see that the magnetic energy is shifted rapidly onto larger scales. For Kolmogorov turbulence the field on the Jeans scale saturates at a density of roughly 3 cm$^{-3}$ and for Burgers at a density of roughly 4 cm$^{-3}$. At the end of dynamo growth on the Jeans scale we have a magnetic field strength of about $10^{-6}$ G throughout the entire inertial range of the turbulence, i.e.\ within the Jeans volume.\\
\begin{figure}
  \centering
  \includegraphics[width=0.45\textwidth]{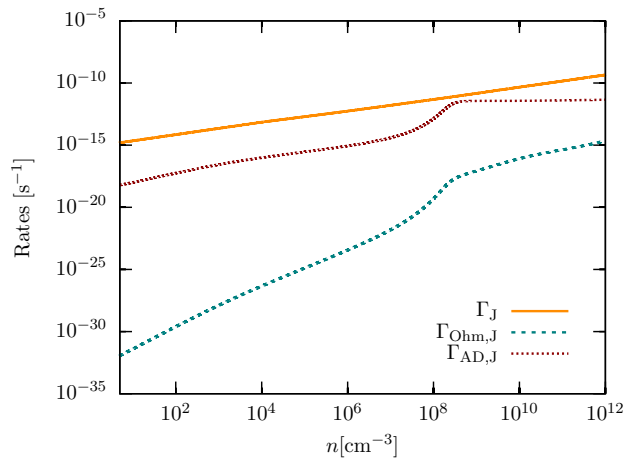}
  \caption{The growth rate on the Jeans scale $\Gamma_\text{J}$ after the dynamo amplification compared to the diffusion rates as a function of the number density. $\Gamma_\text{Ohm,J}$ and $\Gamma_\text{AD,J}$ are the Ohmic and ambipolar diffusion rate, respectively.}
  \label{Figure8}
\end{figure}
After the rapid initial dynamo amplification the only way to amplify the magnetic field on the Jeans scale further is gravitational compression, which leads to $B \propto n^{2/3}$. However, the field has already reached equipartition with the kinetic energy at the end of the dynamo amplification and, thus, increases only with $n^{1/2}$ (see equation \ref{Bmax}). The growth rate of the magnetic field on the Jeans scale $\Gamma_\text{J}$ is then
\begin{equation}
  \Gamma_\text{J} = \frac{1}{n} \frac{\text{d} n}{\text{d} t}.
\end{equation}
In Figure \ref{Figure8} we compare the growth rate $\Gamma_\text{J}$ to the ambipolar and Ohmic diffusion rates on the Jeans scale, $\Gamma_\text{AD,J}$ and $\Gamma_\text{Ohm,J}$. As $\Gamma_\text{J}$ is always larger than the diffusion rates in the shown density range, the magnetic energy on the Jeans scale is not dissipated again during the collapse. At a density of $10^{12}~\text{cm}^{-3}$, we determine with $B \propto n^{1/2}$ a magnetic field strength of 0.4 G.

\subsection{Implications for Numerical Simulations}

Our calculations show that, due to the rather small viscosity and resistivity in primordial gas, the hydrodynamical Reynolds number, the magnetic Reynolds number and the magnetic Prandtl numbers have very high values as long as the magnetic field is not saturated. Such Reynolds numbers are well above what can be reached in numerical simulations, implying that the physical growth rate of the magnetic field largely exceeds the growth rate obtained in numerical simulations. Particularly important here is the fact that the typically unresolved viscous scales are highly relevant for magnetic field amplification even on larger scales. In this sense, numerical simulations can only show the presence of a dynamo, but will typically underestimate the magnetic field amplification rate. This behavior has also been demonstrated in pioneering studies by \citet{SurEtAl2010} and \citet{FederrathEtAl2011a}.\\
On the other hand, our results show that magnetic fields quickly saturate once turbulence forms, and the limiting timescale may thus be the timescale on which turbulence is generated. This is again an issue which can be addressed with numerical simulations, and indeed, simulations for instance by \citet{TurkEtAl2012} convincingly demonstrated the release of turbulence from the gravitational energy during primordial collapse. Overall, such simulations are thus relevant to explore the origin and generation of turbulence, while the strength of the magnetic field should rather be estimated based on the physical growth rates. As a net effect, we therefore expect that the magnetic energy is always close to saturation once turbulence is generated in the halo.

\section{Summary}

We computed the evolution of the magnetic field and its saturation level in typical primordial halos based on the Kazantsev theory of the turbulent dynamo in combination with a detailed description of the physical and chemical processes in zero-metallicity gas. The model is in principle applicable only to magnetic field fluctuations on very small scales. However, when interested in the influence of the field on the overall dynamical evolution of the halo gas, it is most important to understand how saturation occurs on larger scales. To address this problem, we also considered the transport of magnetic energy from the viscous scale to the Jeans scale.\\
Starting with a weak magnetic seed field of $10^{-20}$ G, as can be produced by the Biermann battery, we follow the evolution of magnetic field fluctuations on the viscous scale and found that they are amplified very rapidly on timescales much shorter than the free-fall time. As a consequence, the field saturates almost immediately after the onset of gravitational collapse in the halo. By extrapolating the small-scale magnetic field to larger scales and assuming the peak of the magnetic spectrum shifts on the local eddy timescale, we were able to follow the evolution of the magnetic field strength throughout the full inertial range within the Jeans volume. For typical halo parameters, the dynamo growth of the magnetic energy saturates at a density of roughly $3~\text{cm}^{-3}$ for Kolmogorov turbulence and $4~\text{cm}^{-3}$ for Burgers turbulence. At this point in time the field has a strength of about $10^{-6}$ G. We point out, however, that the field continues to grow in the collapsing gas due to gravitational compression. \\
Our results show that the magnetic energy on small scales, and more importantly also on dynamically important large scales, can grow to very high values. In order to understand the influence of this strong field on the evolution of the halo gas, it is important to know whether the small-scale magnetic field can be transformed into a coherent large-scale field. One way to produce more coherent magnetic structures is by forming disks, which is suggested by \citet{LatifEtAl2011}. Moreover, the saturation behavior of the small-scale dynamo should be explored further in the regime $Pm < 1$, as we have shown that the magnetic Prandtl number is in this regime for high densities.\\
If indeed the processes discussed here can produce dynamically significant fields on large scales, then the magnetic field will influence the star formation process in high-redshift halos. For example, since recent high-resolution simulations indicate that the accretion disks around the very first stars were strongly susceptible to fragmentation \citep{TurkEtAl2009,StacyEtAl2010,ClarkEtAl2011,GreifEtAl2011.1,SmithEtAl2011} it is expected that most primordial stars formed as members of binary or higher-order multiple systems with a wide range of masses rather than being isolated, high-mass stars. From studies of low-mass star formation at present day, however, we know that magnetic fields close to the equipartition value can effectively redistribute angular momentum via a process called magnetic braking \citep{MachidaEtAl2008b,MouschoviasPaleologou1979} and can thereby reduce the fragmentation probability in the disk \citep{HennebelleCiardi2009,PetersEtAl2011,HennebelleEtAl2011, SeifriedEtAl2011}. The correct treatment of magnetic fields in calculations of primordial star formation therefore seems critical to better understand the mass function and multiplicity of metal-free stars.

\acknowledgements
The authors thank Daniele Galli and Wolfram Schmidt for fruitful discussions. Moreover, J.S.~acknowledges financial support by the {\em Deutsche Forschungsgemeinschaft} (DFG) in the {\em Schwerpunktprogramm} SPP 1573 ``Physics of the Interstellar Medium" under grant KL 1358/14-1. D.R.G.S. thanks for funding through the SPP 1573 (project number SCHL~1964/1-1) and the SFB 963/1 {\em Astrophysical Flow Instabilities and Turbulence}. C.F.~acknowledges funding from a Discovery Projects Fellowship of the Australian Research Council (grant~DP110102191). C.F.,~R.B.,~S.G.,~and R.S.K.~acknowledge subsidies from the Baden-W\"urttemberg-Stiftung under research contract P-LS-SPII/18 and from the German Bundesministerium f\"ur Bildung und Forschung via the ASTRONET project STAR FORMAT (grant 05A09VHA). R.S.K also thanks the DFG for financial support via grants KL1358/10 and KL1358/11, as well as via the SFB 881 ``The Milky Way System". R.B. acknowledges funding by the Emmy-Noether grant (DFG) BA~3706 and the DFG via the grands BA 3706/1-1 and BA 3706/3-1.

%\bibliography{Paper2Bib}
%\bibliographystyle{apj}

\end{document}